\documentclass[twoside,11pt]{article}

\usepackage[accepted, specialissue]{melba}
\usepackage{amsmath,amsfonts}
\usepackage{dirtytalk}
\usepackage{booktabs}
\usepackage{multirow}
\usepackage{soul}
\usepackage{caption}
\usepackage{subcaption}
\usepackage{caption}

\melbaid{2024:029}  
\doi{https://doi.org/10.59275/j.melba.2024-682e} 
\melbaauthors{Starck et al.}  

\volume{2}
\firstpageno{2247}  
\melbayear{2024}  
\datesubmitted{03/2024}  
\datepublished{11/2024}  

\melbaspecialissue{Interpretability of Machine Intelligence in Medical Image Computing \\ (iMIMIC) 2023}
\melbaspecialissueeditors{Mauricio Reyes, Jaime Cardoso, Jayashree Kalpathy-Cramer, Nguyen Le Minh, Pedro Abreu, José Amorim, Wilson Silva, Mara Graziani, Amith Kamath}

\ShortHeadings{Atlas-based interpretable age prediction}{Starck et al.}

\title{Atlas-Based Interpretable Age Prediction \\ In Whole-Body MR Images}

\author{\firstname Sophie \surname Starck\orcid{0000-0003-2495-6114} \email sophie.starck@tum.de \\
    \addr Artificial Intelligence in Healthcare and Medicine, School of Computation, Information and Technology, Technical University of Munich, Munich, Germany\AND
\firstname Yadunandan Vivekanand \surname Kini\orcid{0000-0003-1558-0602}  \email ge75yoh@mytum.de \\
    \addr Artificial Intelligence in Healthcare and Medicine, School of Computation, Information and Technology, Technical University of Munich, Munich, Germany \AND
\firstname Jessica J. M. \surname Ritter\orcid{0000-0001-7106-7518}  \email jessica.ritter@tum.de\\
    \addr Institute of Diagnostic and Interventional Radiology, Technical University of Munich, School of Medicine, Munich, Germany \AND
\firstname Rickmer \surname Braren\orcid{0000-0001-6039-6957}  \email rbraren@tum.de\\
    \addr Artificial Intelligence in Healthcare and Medicine, School of Computation, Information and Technology, Technical University of Munich, Munich, Germany \\
    \addr Institute of Diagnostic and Interventional Radiology, Technical University of Munich, School of Medicine, Munich, Germany\\
    \addr German Cancer Consortium (DKTK), Munich partner site, Heidelberg, Germany \AND
\firstname Daniel \surname Rueckert\orcid{0000-0002-5683-5889} \email daniel.rueckert@tum.de\\
    \addr Artificial Intelligence in Healthcare and Medicine, School of Computation, Information and Technology, Technical University of Munich, Munich, Germany \\
    \addr BioMedIA, Department of Computing, Imperial College London, UK \AND
\firstname Tamara T. \surname Mueller\orcid{0000-0002-1818-1036} \email tamara.mueller@tum.de\\
    \addr Artificial Intelligence in Healthcare and Medicine, School of Computation, Information and Technology, Technical University of Munich, Munich, Germany}

\begin{document}

\maketitle

\begin{abstract}
	Age prediction is an important part of medical assessments and research. It can aid in detecting diseases as well as abnormal ageing by highlighting potential discrepancies between chronological and biological age. To improve understanding of age-related changes in various body parts, we investigate the ageing of the human body on a large scale by using whole-body 3D images.
    We utilise the Grad-CAM method to determine the body areas most predictive of a person's age. In order to expand our analysis beyond individual subjects, we employ registration techniques to generate population-wide importance maps that show the most predictive areas in the body for a whole cohort of subjects.
    We show that the investigation of the full 3D volume of the whole body and the population-wide analysis can give important insights into which body parts play the most important roles in predicting a person's age. Our findings reveal three primary areas of interest: the spine, the autochthonous back muscles, and the cardiac region, which exhibits the highest importance. Finally, we investigate differences between subjects that show accelerated and decelerated ageing\footnote{The source code for this work can be found at: \href{https://github.com/starcksophie/atlas_age_pred}{https://github.com/starcksophie/atlas\_age\_pred}\cite{}}. 
\end{abstract}


\section{Introduction}

Deep learning (DL) methods have significantly advanced medical research by delivering insights into normal physiology and disease processes. It can provide imaging-derived biomarkers for non-invasive predictions and support physicians in their work \citep{wang2019deep,piccialli2021survey}. Given the high sensitivity of medical data and the potentially life-altering impact that can result from using DL models for medical diagnoses or interventions, it is important to understand how or why a model reaches its decision.
By inspecting which parts of the input are most relevant to a model's decision, one can examine whether the model actually uses (medically) meaningful information or whether confounders are part of the decision process.

The investigation of ageing, age-related diseases, and the identification of specific areas in the body affected by age have been prominent research areas in medicine. Age shows one of the strongest correlations with the development of diseases and well-being in general \citep{niccoli2012ageing,seale2022making}.
Therefore, acquiring more knowledge about the ageing process can give insights into risk factors or abnormal ageing and serve as an early detection mechanism for several diseases \citep{fayosse2020risk}. 
The utilisation of an accurate age prediction method can aid in (a) establishing a better understanding of the mechanisms of ageing in the human body and (b) finding discrepancies between an individual's chronological and biological age. Chronological age refers to the time elapsed since birth, whereas biological age aims to describe the physiological age, e.g. how the body has aged. There might be deviations between the two, which is often referred to as \textit{accelerated} (biological age $>$ chronological age) or \textit{decelerated} (biological age $<$ chronological age) ageing. This has been investigated extensively for brain age estimation \citep{sajedi2019age} since brain structures are known to change over time \citep{esiri2007ageing,huizinga2018spatio} and be highly correlated with neurodegenerative diseases such as Alzheimer's or Parkinson's disease \citep{luders2016estimating}.
Brain magnetic resonance images (MRIs) are promising modalities to infer the biological brain age of a subject, often with the help of deep learning techniques \citep{sajedi2019age}. 
Age estimation has also been performed on dental data \citep{verma2019dental}, skeleton bones in the body, such as chest radiography \citep{monum2017age}, knee skeletons \citep{maggio2017skeletal}, or hand skeletons \citep{darmawan2015age}. 
Despite significant changes in several abdominal organs and tissues, such as the liver \citep{tajiri2013liver}, bone densities \citep{wishart1995effect}, and the pancreas \citep{meier2007assessment}, whole-body age prediction has so far not been explored in great detail.
However, some works have shown significant advancements in this direction, focusing on the abdominal \cite{le2022using} and whole-body scans \cite{langner2019identifying}.

\begin{figure}[ht!]
     \centering
    \includegraphics[width=1\textwidth]{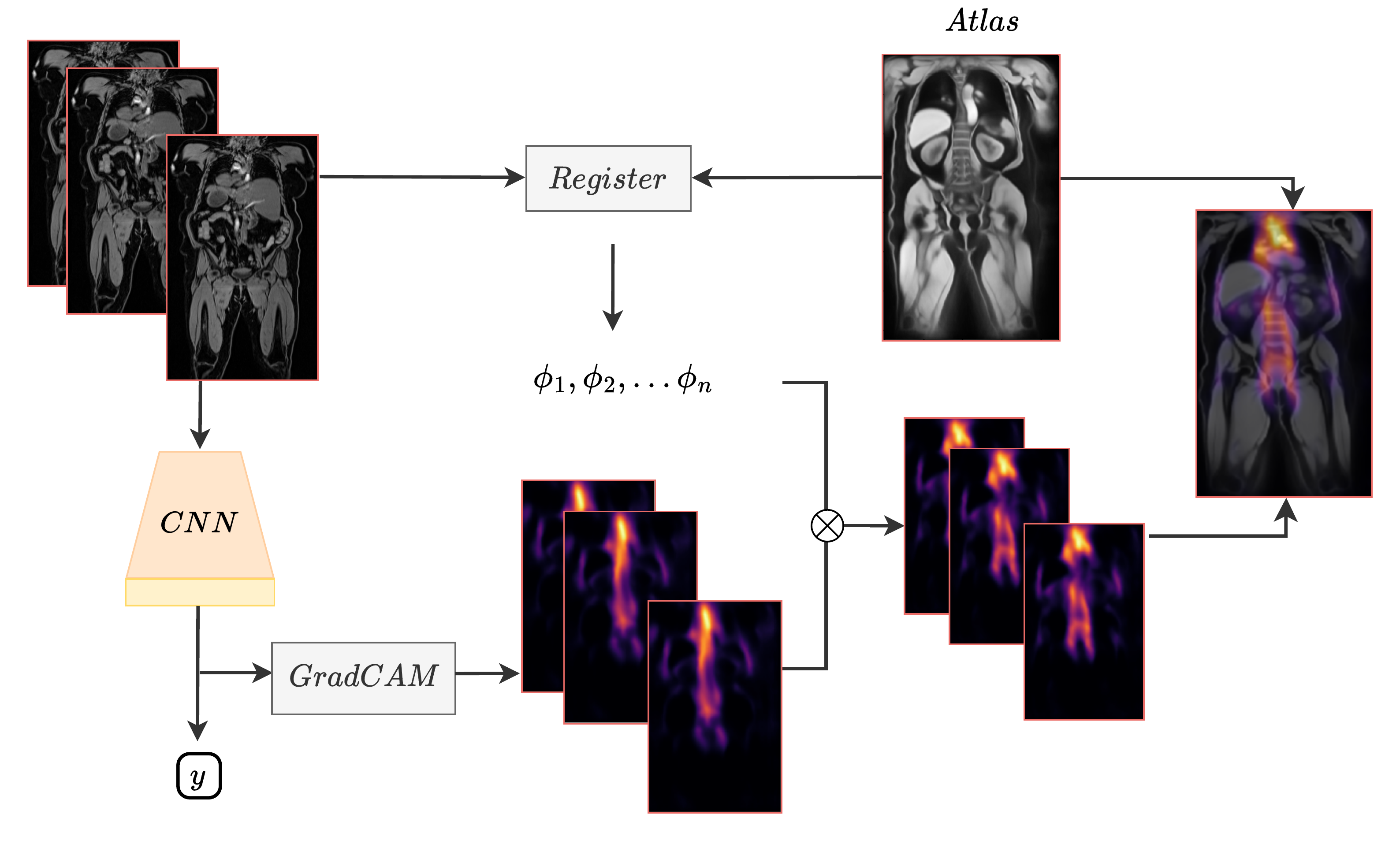}
    \caption{Overview of the pipeline used in this work: First, the CNN is trained to predict age. From the trained model, at inference, Grad-CAM visual explanations are extracted for each subject and then mapped to an atlas before being averaged into a population-wide importance map.}
    \label{fig:archi}
\end{figure}

In this work, we investigate age prediction on the whole body (excluding the brain) to identify which areas show the highest information value about a person's age, utilising the capacity of the whole 3D volumes. Towards this goal, we train a convolutional neural network (CNN) on 3D MR images that cover the full body from neck to knee. Subsequently, we apply Grad-CAM \citep{selvaraju2017grad}, a well-established post-hoc interpretability method for CNNs, to identify areas in the body that are most important to the algorithm's decision-making. 
Since we are specifically interested in the population-wide areas of highest interest for the model, we subsequently register the Grad-CAM results onto a medical atlas to acquire population-wide importance maps. Figure \ref{fig:archi} shows an overview of the pipeline of our work. We identify three main regions of interest in the extracted importance maps: the spine, the autochthonous back muscles, and the heart with its adjacent great vessels like the aorta. Figure \ref{fig:atlas_healthy_f} shows atlas-based importance maps for the healthy female group. We can see that the region along the spine and the area surrounding the heart show the most prominent Grad-CAM activation.

\section{Background and Related Work}
In this section, we summarise relevant background and related works that address interpretability in medical imaging, age prediction, and population-wide studies with medical atlases. 

\subsection{Grad-CAM}
Interpretability methods, such as Grad-CAM can be applied to DL algorithms in order to get a better understanding of the decision-making process of neural networks~\citep{carvalho2019machine}. This is especially important in the medical domain, where critical patient diagnoses might depend on DL predictions, and both physicians and patients might want to understand how or why a model reaches a specific decision. 
One of the most commonly used interpretability methods is gradient-weighted class activation mapping (Grad-CAM) \citep{selvaraju2017grad}.
Grad-CAM utilises the gradient information that flows into a convolutional layer of a CNN and applies global average pooling on these gradients to extract importance values for each input parameter (i.e. image voxel).
Grad-CAM was originally designed for image classification, image captioning, and virtual question-answering tasks \citep{selvaraju2017grad}; it has, since then,  been applied for numerous tasks such as object detection or reinforcement learning  \citep{joo2019visualization,dubost2020weakly}. However, it has been shown that it can also facilitate meaningful interpretations for regression tasks \citep{chen2020adapting}.

Grad-CAM has been used in several applications of DL to medical data \citep{xiao2021visualization,panwar2020deep,daanouni2021automatic} and also specifically in the context of age predictions \citep{langner2019identifying,le2022using, kerber2023deep, bintsi2021voxel, raghu2021deep}.
However, one shortcoming of gradient-based interpretability methods such as Grad-CAM is that the results are subject-specific and do not allow for a population-wide investigation. In the medical context, subject-specific interpretation can be of interest in individual assessments, while a population-level map might hold more generalisable information. In this work, we are interested in population-wide importance maps, which we obtain by using registration methods.

\subsection{Age prediction} 
Ageing is the main risk factor for disease development, and it is an important indicator of a person's overall health \cite{hou2019ageing,niccoli2012ageing}. MR images, in particular, hold great potential in the investigation of the physiological effects of ageing and subsequently identifying diseases.
For instance, deep learning methods have been extensively applied to brain age estimation, achieving a highly accurate age prediction with an error of $2.14$ years \citep{peng2021accurate}.
Age prediction has also been investigated on different body regions such as the teeth \citep{verma2019dental}, the chest \citep{monum2017age}, knees \citep{maggio2017skeletal}, etc.
\cite{le2022using}, have been focusing on abdominal age prediction from liver and pancreas MR images and achieve a mean absolute error (MAE) of $2.94$ years.
The authors also utilise Grad-CAM to highlight the most relevant areas for the model's prediction. However, here, a clear selection of specific regions is difficult as only subject-level areas have been investigated.

The probably most relevant related work to ours is \cite{langner2019identifying}. The authors also perform interpretable age prediction on whole-body images and achieve a performance of $2.49$ years on the UK Biobank \citep{sudlow2015uk}.
Here, constructed projections of the 3D image data into the 2D sagittal and coronal planes are used for training (see Appendix \ref{sec:2d}, Figure \ref{fig:app_2d_comp}). Throughout this paper, we refer to this method as 2.5D since the data is 2D in terms of dimensions, but encompasses information from the entire 3D volume. In \cite{langner2019identifying}, the authors also use Grad-CAM to obtain interpretable maps, indicating which areas of the body are most important for the model to make its prediction. 
They aggregate the resulting saliency maps by co-registering the dataset onto a single representation. Using these projections comes with a major advantage of requiring fewer resources and training time.
However, this method requires significantly more training samples to reach a comparable model performance. We compare our method to their approach and discuss these elements in more detail in Section \ref{sec:results}.
The probably strongest shortcoming of their methods is that the resulting interpretability maps are 2D and highlight regions of the body where all projected slices are overlaid. This makes the actually most meaningful areas indistinguishable from data from other slices.
In order to obtain more precise areas of interest in three dimensions, we here use the full capacity of the 3D volumes.

\subsection{Population-wide Studies and Medical Atlases} Medical imaging is indispensable for medical research and assessment. However, medical images mostly come with high inter-subject variability that can stem from different morphologies or even just different positions in the scanner. Therefore, medical atlases are frequently used to allow for inter-subject or inter-population comparisons. They map several medical images into a common coordinate system, using registration techniques \citep{maintz1998survey}. The registered images are then averaged in order to acquire a template of a specific image modality. This is widely used for brain imaging, where atlases are used to generate an average representation of the human brain \citep{insel2013nih,markram2012human,van2013wu}. Atlas generation on the whole body has been explored considerably less due to a much higher inter-subject variability compared to brain images. However, there are some works focusing on body MR atlas generation that have shown promising applications for these atlases \citep{Sjholm2019AWF,strand2017concept}.

In this work, we utilise conditional atlases generated on a subset of the whole population, split by sex and BMI group (healthy, overweight, obese) \citep{starck2023constructing}.
Consequently, we use six comprehensive whole-body atlases. For each individual, we apply Grad-CAM to generate subject-specific importance maps, which are subsequently aligned with these atlases, yielding population-wide importance maps.

\begin{figure}[ht!]
     \centering
    \includegraphics[width=1\textwidth]{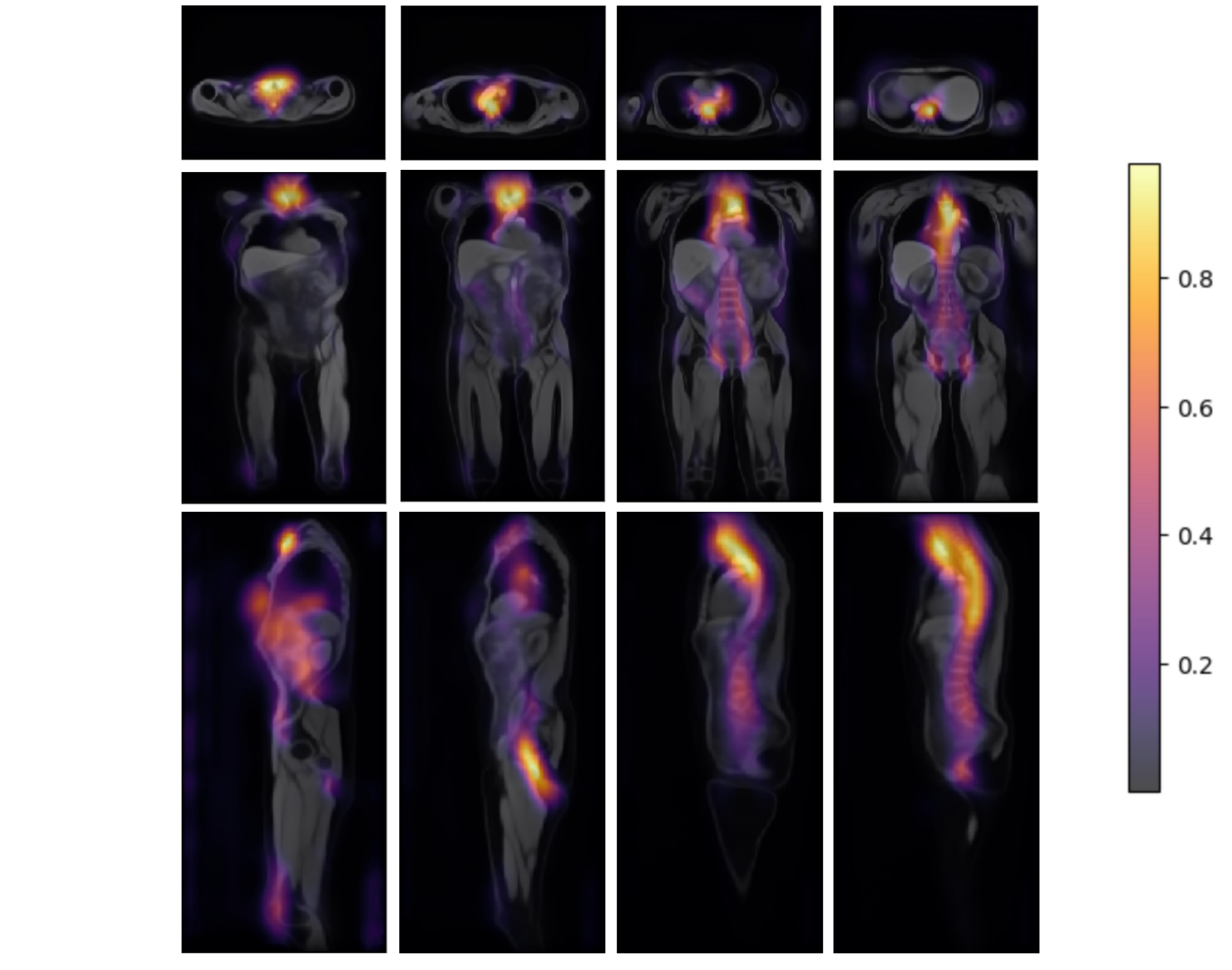}
    \caption{Visualisation of the population-wide Grad-CAM importance maps across several slices (columns) and of different planes (axial, coronal, sagittal) of the healthy female subpopulation, overlaid on the respective atlas.}
    \label{fig:atlas_healthy_f}
\end{figure}
\section{Materials and Methods}
\subsection{Dataset}
The UK Biobank (UKBB) dataset \citep{sudlow2015uk} is a large-scale longitudinal study that has been conducted in the UK since 2006. It contains information from approximately $100\,000$ participants, with a wide range of data such as genetics, biological samples and MR images from the brain, heart, and abdomen. In this work, we utilise the whole-body neck-to-knee MR images acquired with the Dixon technique for internal fat across six stations. We use the water contrast images and stitch the stations together using a publicly available tool \citep{lavdas2019machine}. We select $3120$ subjects with a balanced distribution across age, sex, and BMI. 
$1536$ subjects were used for training, $384$ for validation and $1200$ for testing. The ages range from $46$ to $81$, and the mean age is $63.58$ years. We ensure an equal representation of male and female subjects in all sets.

\subsection{Training Pipeline}
We train a 3D ResNet-18 model \citep{he2016deep,feichtenhofer2019slowfast} from torchvision \citep{NEURIPS2019_9015} with a hidden layer size of $256$.
The training is performed by using adaptive moment estimation (Adam) optimiser \citep{Kingma2014AdamAM} and by minimising the mean absolute error (MAE) of the age predictions. Furthermore, we use a gradient accumulation scheduler which sums and averages the gradients from $32$ consecutive mini-batches to update the model's parameters.
The initial learning rate is $1e-4$, derived from manual tuning and reduced via scheduling when the validation error does not decrease for three epochs.
The model was trained for $100$ epochs, which lasted approximately $48$ hours on an NVIDIA A40 GPU. 

The application of Grad-CAM is independent of the training process. After training the model, we apply Grad-CAM on the third layer of the network using the implementation from \citep{2007.00453}. We apply Grad-CAM at inference and on the test set to evaluate the essential body areas related to age prediction.

Additionally, following similar works in the brain, we apply a statistical bias correction method to the predicted ages to increase accuracy in the prediction and the downstream analysis. Indeed, many age estimation methods suffer from a recurring bias in the overestimation of the age in younger subjects and the underestimation in elders \citep{le2018nonlinear,SMITH2019528}. We use the real age of the validation data as a covariate to predict a bias-corrected age, which we then apply to the test data. An example of the predictions before and after bias correction is available in Appendix \ref{sec:bias}.

\subsection{Registration and Atlas Generation}
We map all subject-level Grad-CAM maps onto an atlas to investigate the important regions for our age prediction model on a population level. Given the high variability of whole-body MR scans, we follow the pipeline proposed in \cite{starck2023constructing} and split all subjects into subgroups depending on their sex and BMI, following three commonly used BMI groups: healthy, overweight, and obese. The registration process is done by first registering all images of a sex and BMI group to the same target subject. We apply two methods: affine and deformable registration. Affine registration refers to a set of rigid transformations such as rotation, translation, shearing, and scaling. These types of transformations allow for a coarse alignment; they do not deform the anatomy of the given subject but only correct the overall position and orientation. The resulting images are then deformed with deformable registration for a more refined registration. This step is more localised and allows for a more detailed alignment. Both registration steps were performed using the publicly available registration tool \texttt{deepali} \citep{deepali}. All parameters are reproduced from \cite{starck2023constructing}. Once all images are registered, the resulting deformation fields are applied to their corresponding activation map, as shown in Figure \ref{fig:archi}. Subsequently, an average map is generated from each subgroup of the dataset which serves as our population-wide importance map. We specifically generate a different overall importance map for the different sex and BMI groups, since there is high anatomical variability between these subgroups. 


\section{Results and Discussion}
\label{sec:results}
We here summarise the results obtained from our experiments, including the age prediction, the extraction of the Grad-CAM importance maps, and the generation of a population-wide importance map.



\begin{table}[ht]
    \centering
    \caption{Summary of the age prediction results by sex and BMI group on the test set. All values are reported MAE scores in years. The \textit{Overall} row reports the MAE of both sexes or all BMI groups, respectively. The score on the whole test set is underlined. The mean prediction refers to a baseline model, that always predicts the mean age of the respective subgroup.}
    \begin{tabular}{lllccccc}
        \toprule
        \textbf{Metric} & \textbf{Category} & \textbf{Sex} & \textbf{Mean Pred.} & \textbf{2.5D} & \textbf{Ours} \\
        \midrule
            \multirow{8}{*}{MAE} & \multirow{2}{*}{Healthy} & F & $7.190$ & $2.485$ & \textbf{2.460} \\
        & & M & $7.672$ & $2.614$ & \textbf{2.425} \\
        \cmidrule{2-6}
        & \multirow{2}{*}{Overweight} & F & $7.485$ & $2.661$ & \textbf{2.525} \\
        & & M & $8.036$ & \textbf{2.327} & $2.623$ \\
       \cmidrule{2-6}
        & \multirow{2}{*}{Obese} & F & $7.045$ & $ 2.863$ & \textbf{2.651} \\
        & & M & $7.550$ &  \textbf{2.669} & $2.720$ \\
        \cmidrule{2-6}
        & Overall & M+F & $7.499$ & $2.613$ & \textbf{2.565}\\
        \midrule
        \midrule
        Nr. training samples & - & - & N/A & $18,384$ & \textbf{1,536} \\
        \midrule
        Runtime/epoch (min) & - & - & N/A & \(\mathbf{1.88 \pm 0.05}\)& $51.33 \pm 1.33$ \\
        Inference/sample (s) & - & - & N/A & \textbf{0.024} & $0.286$ \\
        \bottomrule
    \end{tabular}
    \label{tab:results}
\end{table}

\subsection{Age Prediction}
We evaluate our 3D age prediction model, trained on $1536$ training samples, by randomly selecting $1200$ previously unseen subjects that are approximately equally distributed across all BMI and age groups. Our model achieves a mean absolute error (MAE) of \textbf{2.57} years on this test set. Table \ref{tab:results} summarises the model's performance divided into the same groups that are used for the atlas generation.
As baselines, we utilise a mean prediction for each group (\say{Mean Pred.}) and reproduce the approach from \cite{langner2019identifying} on our test set (\say{2.5D}). We here use a VGG16 and adapt the number of training samples to reach comparable performance.
We can see that the model substantially outperforms the mean prediction (always predicting the mean age of the population), which indicates that it is learning meaningful information. Furthermore, we can see that the model performs best on healthy subjects and performance decreases slightly for the other BMI groups. However, the performance is pretty consistent across all BMI groups and sexes and we do not see a strong bias of the model towards different body composition values or a sex group. In comparison to the 2.5D approach, using the full 3D image volumes leads to slightly better, yet highly comparable, results for all categories apart from the overweight male subjects. 
We also compare the runtime and the number of training samples in Table \ref{tab:results}. In order for the 2.5D approach to achieve comparable performance to the 3D method, it requires approximately $12$ times more training samples. We here utilise $18,380$ training samples to reach a model performance of $2.61$ years on the same test set, compared to $1536$ training samples for the 3D approach. However, its training time per epoch (\say{Runtime/epoch})  and (\say{Inference/sample}) is significantly faster due to the smaller data size. 

A more detailed visualisation of our results is shown in Figure \ref{fig:overview_acc} with a scatter plot of the real age against the predicted one for the whole test set. More detailed plots on all individual groups as well as a table of the results are available in the Appendix, Section \ref{sec:perf}.


\subsection{Extraction of Grad-CAM Maps}
To extract the Grad-CAM maps, we follow the original approach introduced by \cite{selvaraju2017grad}. We extract the importance maps from the inference runs of the $1200$ test subjects ($200$ of each group) and register them to the subgroup atlases to obtain the population-wide attention maps shown in Figure \ref{fig:atlas_healthy_f}. By visual assessment of these individual maps by expert radiologists, we identify three main areas of importance: (1) the spine, (2) the autochthonous muscles of the back, and (3) the heart region, including the myocardium (muscle tissue surrounding the heart) and the aortic arch.
These regions are consistently highlighted over every atlas. Additional highlighted regions comprise the thyroid gland, as we can see in the top left visualisation of Figure \ref{fig:atlas_healthy_f}, the knees (bottom left in Figure \ref{fig:atlas_healthy_f}), the obturator muscles and the abdominal fatty tissue (middle right in Figure \ref{fig:atlas_healthy_f}).
These findings align with related work from \cite{langner2019identifying} as the same regions are consistently highlighted. The use of a 3D model, however, allows for leveraging the entire volume and, therefore, discovering another major region of interest: the spine. More specifically the cervico-thoracic spine region shows strong activations. This could be due to changes in the curvature, i.e. disc degeneration and increased kyphosis of the cervico-thoracic spine, developing with age \citep{yukawa2012age, liu2019effects} \citep{liu2015standard}. Additionally, structural changes in thyroid gland such as calcification and cysts are frequently seen in older patients and tend to increase in size and number with age, which could explain the activations we observe in the thyroid region. Also, degenerative changes of the main axis skeletal joints such as the knees are a frequent finding and scale with aging. The additional regions show lesser importance but are identifiable in all groups.
In short, these findings concur with medical research, as these regions have demonstrated age-related impacts \citep{ignasiak2018effect,paneni2017aging,oei2022imaging}, which provides evidence that these population-wide activation maps hold great potential to investigate which areas in the body contribute most to the model's prediction.

\begin{figure}[ht!]
     \centering
    \includegraphics[width=0.9\textwidth]{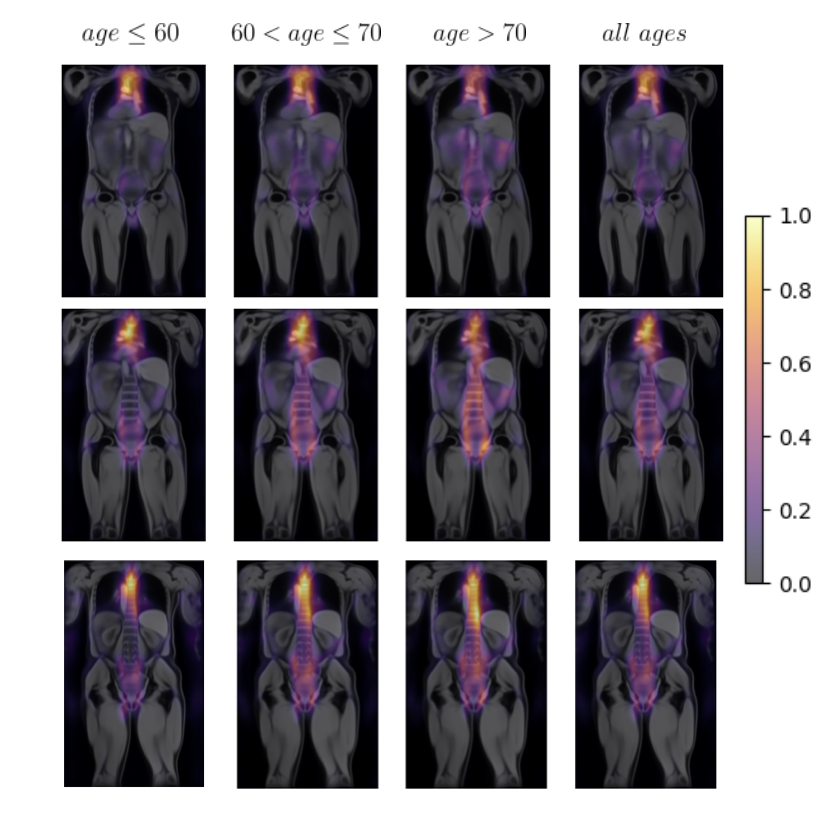}
    \caption{Overview of age group interpretation. All results are shown for the healthy male subgroup. The first column is the population ages lower than 60, the second is between 60 and 70 years old, the third is above 70 and the last is across all groups. One can observe a tendency for stronger activation in the spine area for older subjects.}
    \label{fig:age_groups}
\end{figure}

\subsection{Age-specific Importance Maps}

The groups selected to create the atlases, are chosen with respect to BMI and sex, following \cite{starck2023constructing}. This categorisation aims to model a \textit{global} representation of ageing. 
Our results align with the current understanding of how and where ageing is happening in the human body and we did not observe any notable differences between the different BMI and sex groups. 
However, we are also interested in whether there are any differences in the importance maps between different age groups.
We explore this by generating population-wide importance maps for three different age groups: subjects below $60$ years old, between $60$ and $70$ years old and above $70$ years old (approximately 70 samples per group).
Figure \ref{fig:age_groups} visualises the importance maps for these three groups within the healthy male group. Here, we observe a noteworthy change across age groups, highlighting differences in characteristic regions of importance in each cohort. 
In particular, we note that the focus on the spine increases with age. Ageing comes indeed with various spine-related disorders such as degenerative scoliosis or osteoporosis ~\citep{fehlings2015aging}, which potentially guides the model's predictions for subjects in older age categories. 
Another finding is that the significance of the autochthonous back muscle region appears to grow with age, potentially due to the increased prevalence of sarcopenia in older individuals~\citep{tournadre2019sarcopenia}, a muscle disorder associated with ageing inducing lower muscle mass.

This analysis might indicate different ageing patterns over age, enabling the detection of age-related features specific to a subpopulation, e.g. frailty, and diseases.
We see strong potential in the analysis of further subgroups and how ageing potentially impacts different regions of the body for specific diseases such as cardiovascular disorders or diabetes.

\begin{figure}[ht!]
    \centering
    \includegraphics[width=1\textwidth]{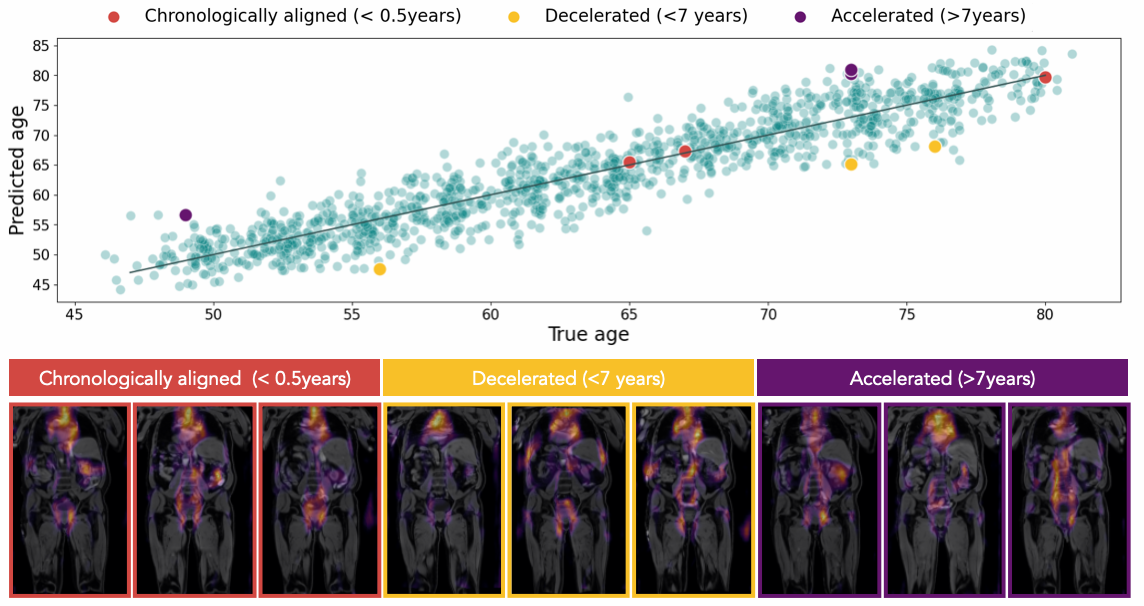}
    \caption{Overview of the model predictions (y-axis) versus actual age (x-axis) for all groups. Below is a visualisation of importance maps for three groups in the overweight male category: (a) subjects where the prediction was near perfect ($\text{MAE} < 0.5$) in red, (b) subjects where the age was underestimated (decelerated) in yellow, and (c) subject where the age was overestimated (accelerated) in purple. These examples are highlighted on the above scatter plot with their respective colours.}
    \label{fig:overview_acc}
\end{figure}

\subsection{Chronological vs. Predicted Age Gaps}
Individual importance maps provide insights into the regions that informed the model to make its prediction for a subject.
We have leveraged this information to derive global features on a population level and have compared different groups of interest.
These general maps emphasize strong predictors of ageing and remove any diffuse noise derived from Grad-CAM on the level of individual samples. A summary of the model performance is visible in Figure \ref{fig:overview_acc}. The scatter plot of the predicted age versus the chronological age highlights the error as the distance to the identity function, i.e. the more distant a point is from the identity (black line), the higher the error. In some cases, the predicted age of a subject can differ significantly from their chronological age. This can be attributed to two reasons: either (a) the model fails to predict the actual age of a subject or (b) the subject exhibits signs of accelerated or decelerated ageing.

We investigate this by visualising the individual importance maps (Figure \ref{fig:overview_acc}) for strong mis-predictions to assess whether model failure is visible in them.
We visualise three examples for the most heavily accelerated (purple) and decelerated (yellow) data alongside \say{near perfect} predictions (red) for the overweight male group (Figure \ref{fig:overview_acc}). The variability between these maps is quite high and aside from marginal noise in the abdominal region, no consistent deviation from the atlas is apparent in the maps. Given the fact that the importance maps do not show any sign that these are mis-predicted, we conclude that either Grad-CAM is inadequate to reflect mis-predictions in this setting, or the model's decision is informed by age-related features and the subjects are indeed accelerated agers.

Since consistent deviations from the atlas are difficult to detect on an individual basis, we aggregate the importance maps for the subjects where we observe accelerated and decelerated ageing. Comparing these maps, we do observe a difference for the accelerated age group; the spine and autochthonous muscles show stronger activations. This indicates that there might be physiological differences for accelerated ageing subjects compared to normal and decelerated ageing.
However, validating this finding is challenging for several reasons. Firstly, the model was not exclusively trained on healthy data, which is typically done to ensure that the training subjects' chronological age matches their biological age \citep{tian2023heterogeneous}. To conduct a more thorough analysis of accelerated and decelerated ageing, a specifically designed training regime might be necessary. Additionally, the importance maps generated by Grad-CAM are not highly precise, and the strong signals from other parts of the dataset, such as the heart and spine, might obscure the detection of accelerated features. The model tends to utilise the simplest most predictive features in the input data, which is why secondary regions that still contain important information about the age of a subject might be ignored by the model.
Therefore, further investigation would be needed to claim that the importance maps highlight accelerated or decelerated ageing features.


\begin{figure}[ht!]
    \begin{subfigure}{.3\textwidth}
      \centering
      \includegraphics[width=.9\linewidth]{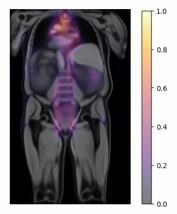}
      \caption{Chronologically aligned}
      \label{fig:ood_chron}
    \end{subfigure}
    \begin{subfigure}{.3\textwidth}
      \centering
      \includegraphics[width=.9\linewidth]{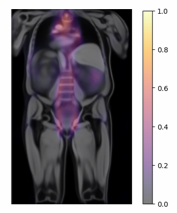}
      \caption{Accelerated}
      \label{fig:ood_acc}
    \end{subfigure}
    \begin{subfigure}{.3\textwidth}
      \centering
      \includegraphics[width=.9\linewidth]{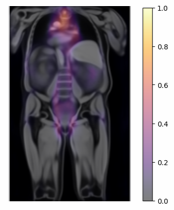}
      \caption{Decelerated}
      \label{fig:ood_dec}
    \end{subfigure}
\caption{Visualisation of group-wise importance maps in the overweight male category for predictions that are aligned, accelerated and decelerated with respect to the chronological age. (\ref{fig:ood_chron}) is the chronologically align atlas, (\ref{fig:ood_acc}) the accelerated one and (\ref{fig:ood_dec}) the decelerated one. We observe stronger activation in the spine for the accelerated age group compared to the decelerated age group.}
\label{fig:ood_atlas}
\end{figure}

\section{Conclusion and Future Work}
In this work, we investigate which areas in the body contribute most to our whole-body MRI age predictor.
We train a 3D ResNet-18 model on $1536$ neck-to-knee MR images from the UK Biobank \citep{sudlow2015uk}. Our model performs whole-body age prediction with a mean absolute error of $2.57$ years on the test set after bias correction.
To investigate the most predictive parts of the body, we apply Grad-CAM to the gradients derived from each test subject. However, these importance maps are subject-specific and do not easily generalise to the whole population. We address this by registering all importance maps into the same coordinate space, aggregating them, and overlaying them with an atlas of whole-body MR images. We here use six distinct groups in the population, based on their sex and BMI.
The aggregated importance maps for all subjects highlight three main regions of interest: the spine, the cardiac region, and the autochthonous back muscles. Despite mapping individual importance maps to the population atlas and across various groups, these areas stay consistent.
We also investigate differences in importance maps across age groups and find that the model places a stronger focus on the spine and autochthonous back muscles in older subjects. This may be due to older individuals being more likely to suffer from spine-related issues, influencing the model's predictions.
In all experiments, the highlighted areas of the importance maps align with medical knowledge and previous studies on ageing. This alignment is encouraging, suggesting that examining the importance maps for pathological groups might reveal new insights into the association between DL-based age prediction and specific pathologies. 
Additionally, we analyse the generated importance maps for subjects that show where the deviation between model prediction and chronological age is high. We hereby distinguish between over- and under-estimated age predictions and compare the averaged importance maps of these two groups. 
We do not observe any significant changes in individual importance maps for these subjects. However, aggregating them
demonstrated slightly stronger activations in the spine for the accelerated ageing group. This could indicate physiological differences in accelerated ageing subjects compared to those with normal and decelerated ageing. However, further investigation would be required to confirm this finding, which we consider an interesting direction for future work.


We envision several other interesting directions of future work to further investigate highly relevant areas in the human body for DL-based age predictors. The here-generated importance maps primarily highlight specific regions of interest in the body, potentially neglecting other areas that may contain valuable information about ageing. An approach to address this, and focus on secondary body regions would be to mask the input images during training, either randomly or by deliberately omitting the most predictive areas (such as the spine, aortic arch, and back muscles). This would steer the model towards different features for its predictions, potentially discovering valuable medical insights. 
Furthermore, we identify the opportunity to extract even more qualitative importance maps as an interesting next step. We aim to further investigate different interpretability methods, such as perturbation-based methods \citep{zeiler2014visualizing} or attention-based models, such as Vision Transformers \citep{dosovitskiy2020image} and compare their results to the here utilised Grad-CAM method.
Moreover, we intend to implement our method on comparable datasets like the German National Cohort \citep{bamberg2015whole} or in-house hospital data, and therefore a wider age range than the one represented in the UK Biobank, to validate the broader applicability of these findings.
We believe that the here showcased move from individual insights into the decision-making process of DL methods to a population-level has great potential to further investigate the interplay between DL and medical research.


\acks{SS and TTM were supported by the ERC (Deep4MI - 884622). This work has been conducted under the UK Biobank application 87802. SS has furthermore been supported by the Federal Ministry of Education and Research (BMBF). RB was funded by the Federal Ministry of Education and Research (BMBF, Grant Nr. 01ZZ2315B and 01KX2021), the Bavarian Cancer Research Center (BZKF, Lighthouse AI and Bioinformatics) and the German Cancer Consortium (DKTK, Joint Imaging Platform).}

%
\ethics{The work follows appropriate ethical standards in conducting research and writing the manuscript, following all applicable laws and regulations regarding the treatment of animals or human subjects.}

\coi{We declare we don't have conflicts of interest.}

\data{The UK Biobank dataset is available on \href{www.ukbiobank.ac.uk}{www.ukbiobank.ac.uk}, upon registration. For this study, permission to access and analyse the UK Biobank data was approved under the application 87802
} 

\bibliography{literature}


\clearpage
\appendix
\section{Atlases}
\begin{figure}[ht!]
     \centering
    \includegraphics[width=\textwidth]{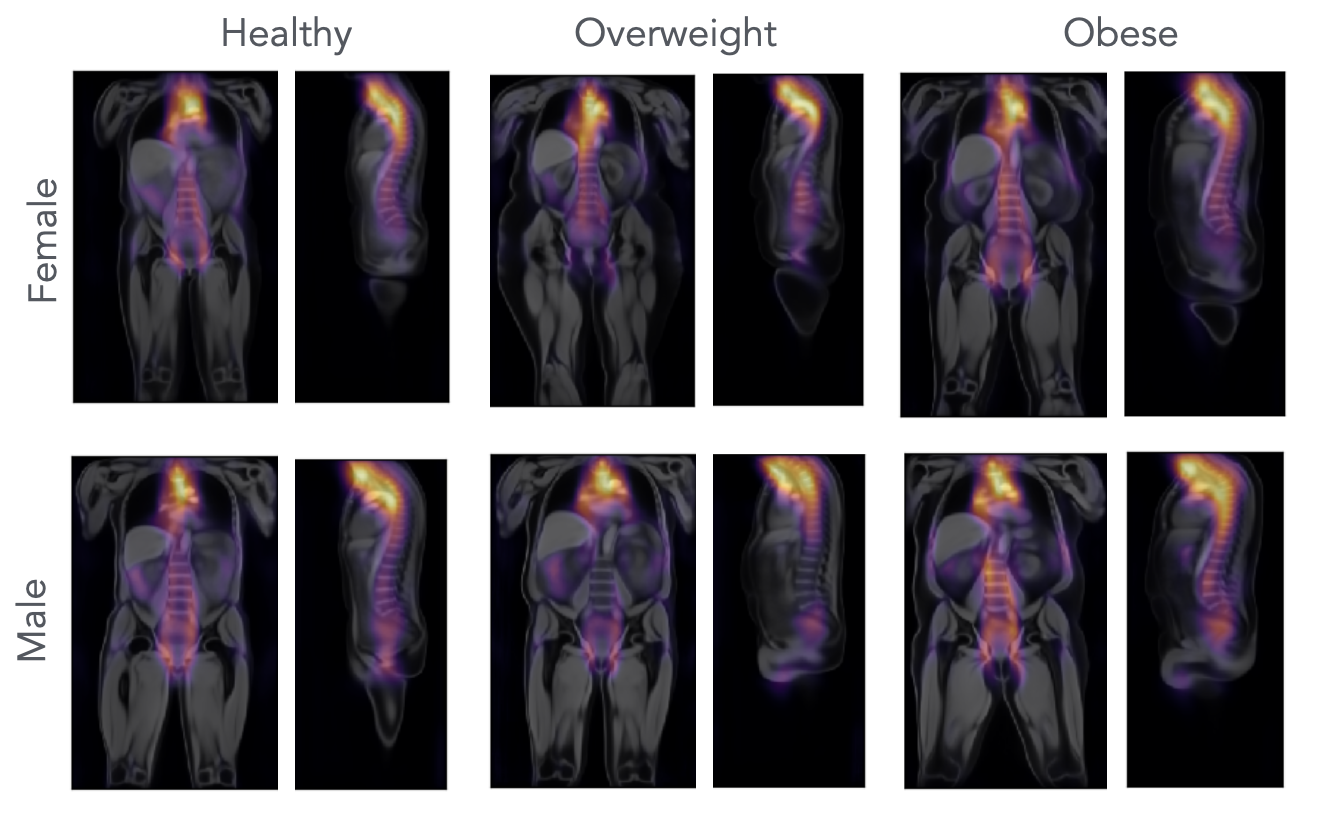}
    \caption{Overview of the population-wide Grad-CAM activation maps across all categories, overlaid on the respective atlas.}
    \label{fig:appendix_atlases}
\end{figure}

Figure \ref{fig:appendix_atlases} shows the population-wide Grad-CAM maps for both sexes and all BMI groups. We here selected one coronal and one sagittal slice to show the main activations in both views. All activation maps highlight similar regions as discussed in the main part of the paper and we do not observe a meaningful difference between sexes or BMI groups.

\newpage

\section{Bias correction}
\begin{figure}[ht!]
    \centering
    \includegraphics[width=1\textwidth]{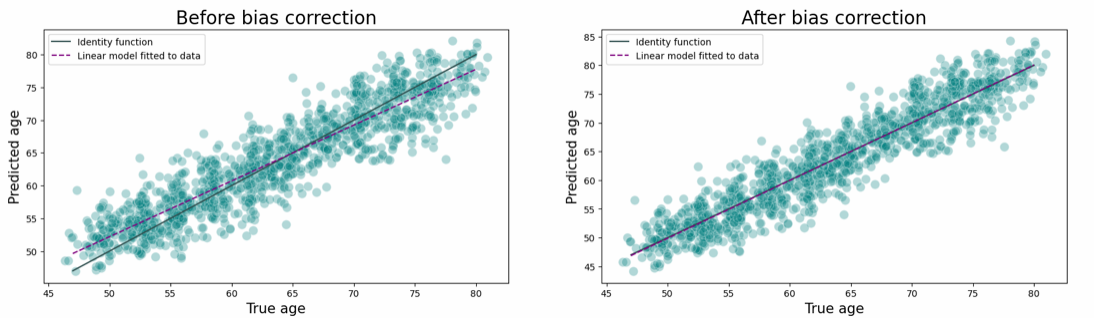}
    \caption{Visualisation of the error on the test set before and after bias correction. The left scatter plot shows the data before correction and the bias is clearly visible and the right plot shows the performance after correction.}
    \label{fig:bias}
\end{figure}

Similar to many other age prediction works \cite{tian2023heterogeneous,le2018nonlinear,SMITH2019528}, we apply a bias correction to the initial age predictions of our model. The scatter plots of the predictions before and after bias correction are visualised in Figure \ref{fig:bias}.

\label{sec:bias}
\section{Summary performance}
\label{sec:perf}
\begin{figure}[ht!]
    \centering
    \includegraphics[width=1\textwidth]{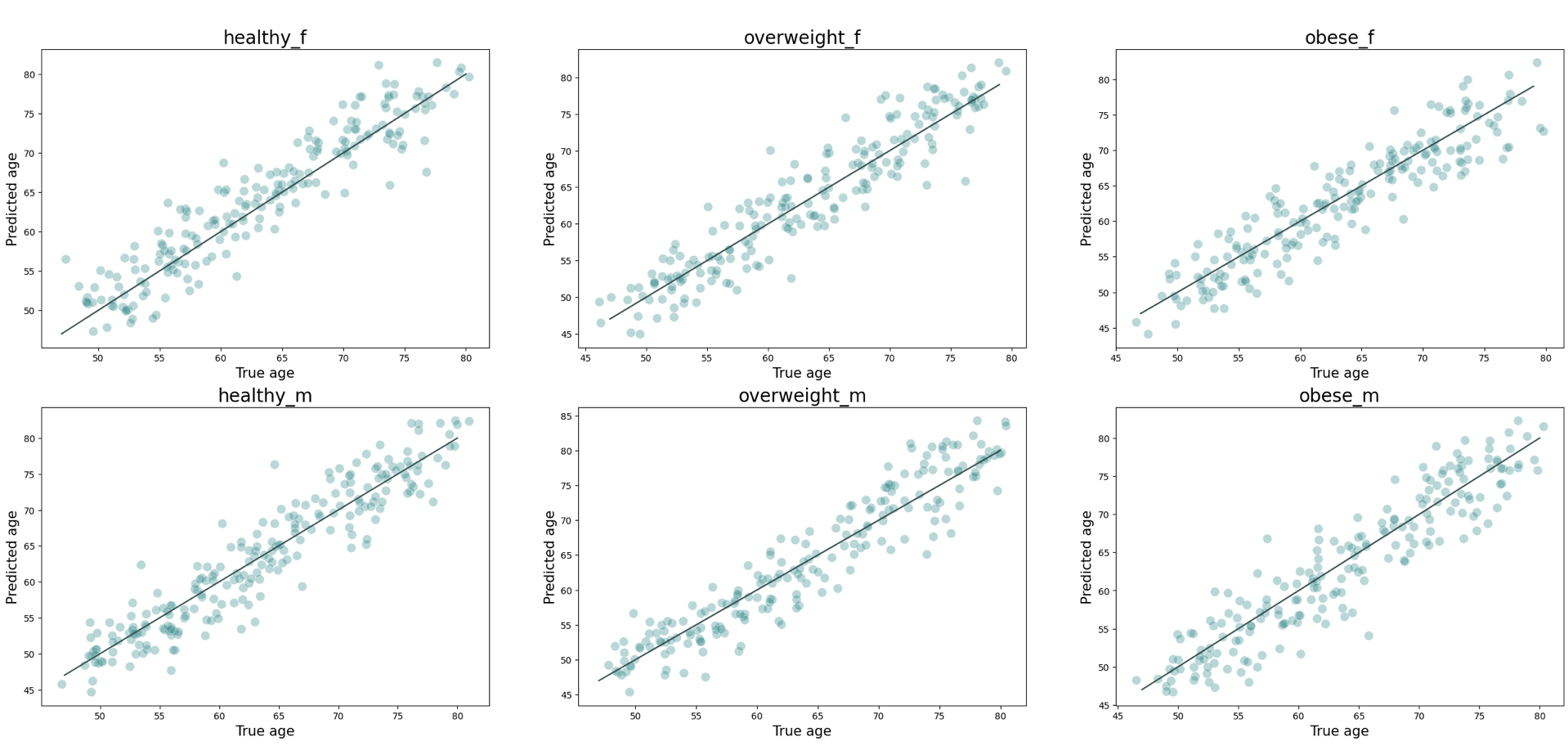}
    \caption{Overview of the bias-corrected age prediction of the model versus actual age for all groups. The top row shows the performance for the female group and the second for males for respectively the healthy, overweight and obese groups}
    \label{fig:truevspredall}
\end{figure}

Figure \ref{fig:truevspredall} shows the scatter plots of the chronological age against the bias-corrected predicted age for all subgroups. These results are furthermore summarised in Table \ref{tab:results} in the main manuscript listing the respective mean absolute errors.

\section{Comparison to 2.5D}
\begin{figure}[ht!]
    \begin{subfigure}{.5\textwidth}
      \centering
      \includegraphics[width=.8\linewidth]{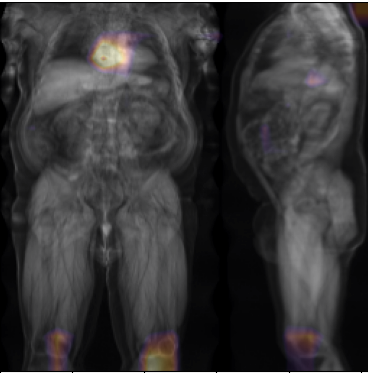}  
    \end{subfigure}
    \begin{subfigure}{.5\textwidth}
      \centering
      \includegraphics[width=.8\linewidth]{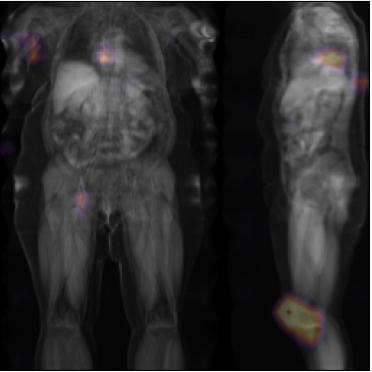}  
    \end{subfigure}
\caption{Visualisation of two example test subjects and their interpretable Grad-CAM maps for the 2.5D approach. Here the heart region and the knee area seem to contribute mostly to the model's prediction. However, it is difficult to identify which specific regions in 3D space are the most relevant, as they are all projected to the same plane.}
\label{fig:app_2d_comp}
\end{figure}

We compare our approach to a less resource-hungry one that works on projected images from the 3D volume, which we call 2.5D. We follow the approach from \cite{langner2019identifying}. Two example 2.5D images and their corresponding importance maps are visualised in Figure \ref{fig:app_2d_comp}. We achieve very similar attention maps as reported in their work, however, highlight that the as important considered regions are difficult to assign to specific regions in 3D space due to the nature of the projected images. 

\label{sec:2d}

\end{document}